\documentclass[12pt,a4paper]{article} 
\usepackage{amsfonts}
\vspace{20mm}

\let\ddpp\mathbb 
\def\R{\ddpp R}     
\def\S{\ddpp S}     

\begin{document}

\title{\LARGE {\bf A Geometric Algorithm to construct new solitons in the $O(3)$ Nonlinear Sigma Model}}
\author{{\bf Manuel Barros} \thanks{e-mail:mbarros@ugr.es}\\
Departamento de Geometr\'\i{}a y Topolog\'\i{}a\\Universidad de Granada\\18071, Granada, Spain}
\date{}
\maketitle

\begin{abstract} The $O(3)$ nonlinear sigma model with boundary, in dimension two, is considered. An algorithm to
determine all its soliton solutions that preserve a rotational symmetry in the boundary is
exhibited. This nonlinear problem is reduced to that of clamped elastica in a hyperbolic plane.
These solutions carry topological charges that can be holographically determined from the boundary
conditions. As a limiting case, we give a wide
family of new soliton solutions in the free $O(3)$ nonlinear sigma model. \\

\end{abstract}



\noindent
\section{Introduction}
It is well known the existence of a close link between the differential geometry of surfaces in
Euclidean three-space and a wide variety of non-linear phenomena in physics and mathematics. This
setting including problems in geometric analysis, continuum mechanics, sigma models, string
theories, theory of biological membranes etc. (see as an example
\cite{Arroyo-Barros-Garay1,Biscari,Concus-Lancaster} and references therein). Many of those
problems are related with functionals of the type
\[ {\cal B}(M)=\int_M F(dN)dA, \]
where $M$ is a surface in $\R^3$ with Gauss map $N$, $F$ is a certain function and $dA$ is the
element of area, on $M$, of the induced metric from the Euclidean  one.

Now, people are interested in variational problems related with these functionals. In particular,
in minimizing them in a given class, ${\cal M}$, of surfaces satisfying certain constraints either
topological (free Willmore, free $O(3)$ nonlinear sigma model, genus one biological membranes,
constant mean curvature surfaces etc.) or/and boundary conditions (Plateau, Willmore with boundary,
$O(3)$ nonlinear sigma model with boundary etc.)

On the other hand, the $O(3)$ nonlinear sigma model (NSM) is ubiquitous in physics (see for example
\cite{Cavalcante, Tsurumaru} and references therein). It is used in a wide ranging of fields from
condensed matter physics, (see for example \cite{Belavin-Polyakov,Laughlin}), to high energy
physics (see for instance \cite{Cecilia1,Cecilia2}). The NSM, in dimension two, plays an important
role in string theories where the model description is applicable. However, it has its own interest
in connection with differential geometry, for example, it naturally leads to the appearance of
partially integrable almost product structures, \cite{Cecilia1, Cecilia2}.

\vspace{2 mm}

\noindent In this note we deal with the two-dimensional NSM, in general with non trivial boundary
values. We obtain all the solutions which preserve a considered rotational symmetry in the boundary
conditions. To do it, we exhibit an algorithm with two main phases. In the first, that could be
called the reduction phase and culminates in the theorem 2, we phase out the search of symmetric
solitons in the NSM with rotational symmetric boundary to that of clamped elasticae in a hyperbolic
plane. In a second phase, we give the description of the whole space of clamped elasticae in the
hyperbolic plane, \cite{Langer-Singer1}. The topological charges, that these solitons carry, can be
holographically determined from the boundary conditions (see the corollary).

\vspace{2 mm}

\noindent As a limiting case, we discover an ample family of new soliton solutions in the free NSM.
In particular, the qualitative behaviour of those solitons obtained by rotation of wavelike
elastica, can be explained in the following points:
\begin{enumerate}
\item
The solutions are topologically twice punctured spheres obtained as surfaces of revolution whose
profile curves are wavelike elasticae in the hyperbolic plane.
\item
They oscillate along meridians of twice punctured round spheres.
\item
The topological charge of these solitons vanishes identically, even though they can be completed to
obtain Zoll's surfaces that present two singularities.
\end{enumerate}

\section{The boundary free NSM}
The elementary fields in the two dimension, boundary free, NSM are $\R^3$-valued unit vector fields
on surfaces without boundary. An interesting approach to study this model, in connection with the
differential geometry of surfaces in three dimensional Euclidean space, was considered in
\cite{Ody}. In this context, one identifies the unit normal vector field, or, more correctly, the
Gauss map of a surface in $\R^3$ with the dynamical variable of the NSM. To be precise, let $M$ be
a surface and denote by $I(M,\R^3)$ the space of immersions of $M$ in the Euclidean space,
$(\R^3,g=<,>)$. For any $\phi\in I(M,\R^3)$, we have its Gauss map, $N_{\phi}:M\rightarrow\S^2$.
Therefore, $dN_{\phi}$ denotes the shape operator of $\phi$. Now, the field configuration of the
NSM can be identified with $I(M,\R^3)$ and the Lagrangian that governs the dynamics of the model,
${\cal D}:I(M,\R^3)\rightarrow\R$, measures the total energy of the Gauss mappings, that is,
\[ {\cal D}(\phi)=\int_M \mid dN_{\phi}\mid^2dA_{\phi}, \]
where $dA_{\phi}$ denotes the element of area of $(M,\phi^* g)$. The known solutions to the
boundary free NSM can be obtained from the theory of surfaces with constant mean curvature. For
example, the solitons discovered by A.A.Belavin and A.M.Polyakov, \cite{Belavin-Polyakov},
correspond with those surfaces whose Gauss maps are conformal (round spheres and minimal surfaces).
Also, the solutions given by S.Purkait and D.Ray, \cite{Purkait-Ray}, are induced by the family of
constant mean curvature helicoids studied by M.P.Do Carmo and M.Dajzer, \cite{DoCarmo-Dajzer}.

\vspace{2 mm}

\noindent We denote by $H_{\phi}$ the mean curvature function of $\phi\in I(M,\R^3)$ and put
$G_{\phi}=det(dN_{\phi})$ to name the Gaussian curvature of $(M,\phi^*(g))$. The following
relationship is classical
\[ \mid dN_{\phi}\mid^2 = 4H_{\phi}^2-2G_{\phi}. \]

\noindent When $M$ is assumed to be compact then, one can use the Gauss-Bonnet theorem to obtain

\vspace{2 mm}

\noindent {\bf Theorem 1} {\it Let $M$ be a compact surface then, $\phi\in I(M,\R^3)$ is a soliton
of the NSM if and only if $(M,\phi)$ is a Willmore surface, that is $\phi$ a critical point of the
action ${\cal W}:I(M,\R^3)\rightarrow\R$, given by
\[ {\cal W}(\phi)=\int_M H_{\phi}^2 dA_{\phi}. \]}

\vspace{2 mm}

\noindent This result can be used to generate wide families of compact soliton solutions of the
free NSM (combine, for example with the classes of Willmore surfaces obtained in
\cite{Arroyo-Barros-Garay,Barros,Langer-Singer2,Lawson,Pinkall}).

\section{The NSM with boundary}

The Willmore functional and the NSM action, both in the free boundary approach, are particular
cases of functional ${\cal B}(\phi)=\int_M F(dN_{\phi})dA_{\phi}$. From now on, we will deal with
the following problem, which can be considered as the NSM with boundary conditions. It has been
widely considered along the literature (see for example \cite{Anzellotti1,Anzellotti2} and
references therein).

\vspace{2 mm}

\noindent {\it THE FIRST ORDER BOUNDARY CONDITIONS. We consider the boundary conditions
$(\Gamma,N_o)$, where $\Gamma=\{\gamma_1,\gamma_2,\cdots,\gamma_n\}$ is a finite set of regular
closed curves in $\R^3$ with $\gamma_{i}\cap\gamma_{j}=\emptyset$ if $i\neq j$, given $x\in\Gamma$,
we put $\Gamma^{\prime}(x)=\gamma_{j}^{\prime}(x)$ if $x\in\gamma_{j}$ to denote the tangent vector
field of $\Gamma$. Now, $N_o$ is a unit normal vector field along $\Gamma$ and such that
$<N_o(x),\Gamma^{\prime}(x)>=0$, $\forall x\in\Gamma$. In this setting, we have a vector field,
$\nu$ along $\Gamma$ determined by $\Gamma^{\prime}(x)\wedge\nu(x)=N_o(x)$, $\forall x\in\Gamma$.}

\vspace{2 mm}

\noindent {\it THE BOUNDARY VALUE PROBLEM. Let $M$ be a differentiable surface with boundary
$\partial M=c_1\cup c_2\cdots\cup c_n$. We denote by $I^{D}_{\Gamma}(M,\R^3)$ the space of
immersions, $\phi:M\rightarrow\R^3$, that satisfy the following boundary conditions
\begin{enumerate}
\item $\phi(\partial M)=\Gamma$, or $\phi(c_j)=\gamma_j$, $1\leq j\leq n$, and
\item $d\phi_{p} (T_p M)$ is orthogonal to $N_o(\phi(p))$, $\forall p\in\partial M$.
\end{enumerate}

\noindent Now the problem is to study the dynamics of the model ${\cal
D}:I^{D}_{\Gamma}(M,\R^3)\rightarrow\R$ given by

\[{\cal D}(\phi)=\int_M\mid dN_{\phi}\mid^2dA_{\phi}. \]
Roughly speaking, if we identify each immersion $\phi\in I^{D}_{\Gamma}(M,\R^3)$ with its graph
$\phi(M)$, viewed as a surface with boundary in $\R^3$, then we propose the study of the Lagrangian
${\cal D}$ in the class of surfaces with the same boundary and with the same Gauss map along the
common boundary.}

\vspace{2 mm}

The amazing fact is that the NSM with boundary and the Willmore problem with boundary,
\cite{Weiner}, are equivalent. A result similar to that obtained in the theorem 1 for closed
(compact and boundary free) case. To prove this claim we observe that, via the Gauss-Bonnet
formula, the NSM action can be written as

\[ {\cal D}(\phi)=\int_{M}\mid dN_{\phi}\mid^2dA_{\phi}=4\int_{M}H_{\phi}^2dA_{\phi}+2\sum_{i=1}^n \int_{\gamma_i}\kappa_i^{\phi} ds, \]
where $\kappa_i^{\phi}$ stands for the curvature function of $\gamma_i$ in $\phi(M)$ endowed with
the $g$-induced metric. \noindent However, the functional ${\cal
L}:I_{\Gamma}^D(M,\R^3)\rightarrow\R$ given by
\[ {\cal L}(\phi)=\int_{\phi(\partial M)} \kappa^{\phi} ds=\sum_{i=1}^n \int_{\gamma_i}\kappa_i^{\phi} ds, \]
is constant on the whole $I_{\Gamma}^D(M,\R^3)$. In fact, this follows from the stated boundary
conditions which imply that the curvature, $\kappa^{\phi}$, of $\phi(\partial M)=\Gamma$ in $\phi
(M)$, endowed with the $g$-induced metric, actually does not depend on $\phi\in
I_{\Gamma}^D(M,\R^3)$ because $\kappa^{\phi}$ comes from the projection of the boundary
acceleration, $\Gamma^{\prime\prime}$, on $d\phi_p(T_p M)$, which is the tangent plane of each
$\phi(M)$ because all the immersions have the same Gauss map along the common boundary $\Gamma$.

\noindent Consequently, the NSM with boundary is equivalent to the following Willmore problem with
boundary, \cite{Weiner}

\[ {\cal W}(\phi)=\int_{M}H^2_{\phi}dA_{\phi}+\int_{\phi(\partial M)}\kappa^{\phi} ds. \]
In particular, the class of soliton solutions in the NSM with boundary coincides with the class of
Willmore soliton surfaces with boundary.

\section{The NSM with rotational symmetry in the boundary}
The above boundary conditions are invariant under a rotational group of symmetries, $G=SO(2)$, with
axis $L$ if and only if: (1) The boundary, $\Gamma$ is made up of two circles,
$\{\gamma_1,\gamma_2\}$, in parallel planes orthogonal to $L$. (2) Moreover, the angle that $\nu$
(equivalently $N_o$) makes with the axis, $L$, is constant along each component of $\Gamma$. We
denote by $\theta_i\in[0,\pi]$, the angle that $\nu$ makes with $L$ along $\gamma_i$, $1\leq i\leq
2$. These are nothing but the angles that the tangent plane, $d\phi_p(T_pM)$, along each admissible
immersion makes with the axis. Therefore, the admissible immersions, $\phi\in
I^{D}_{\Gamma}(M,\R^3)$, that have the same boundary and the same Gauss map along the common
boundary, now also satisfy that $N_{\phi}(\Gamma)$ lie in two circles of $\S^2$.

\vspace{2 mm}

\noindent {\bf Reduction of symmetry.} We consider the cylinder $M=\S^1\times [a_1,a_2]$, then
$\partial M=c_1\cup c_2$, where $c_i=\S^1\times \{a_i\}$, $1\leq i\leq 2$. In this case, $G$ acts
naturally on $I_{\Gamma}^D(M,\R^3)$ by $(f,\phi)\mapsto f\circ\phi$, $\forall f\in G$. Furthermore,
the NSM Lagrangian ${\cal D}:I_{\Gamma}^D(M,\R^3)\rightarrow\R$ is obviously $G$-invariant, i.e.,
${\cal D}(f\circ\phi)={\cal D}(\phi)$, $\forall f\in G$ and $\forall\phi\in I_{\Gamma}^D(M,\R^3)$.

The principle of symmetric criticality, \cite{Palais}, can be applied in this setting. It works in
the following sense. Let $\Sigma_{G}$ be the space of immersions, $\phi\in I_{\Gamma}^D(M,\R^3)$,
which are $G$-invariant, that is $f\circ\phi=\phi$, $\forall f\in G$, we will refer these
immersions as symmetric points. Then, a symmetric point, $\phi\in\Sigma_G$, is a solution in the
NSM with rotational symmetric boundary if and only if it is a critical point of ${\cal
D}:\Sigma_G\rightarrow\R$. In other words, the $G$-invariant solutions of the field equations
coincide with the solutions of the $G$-reduced field equations.

To compute this restriction, first, we need to identify the space $\Sigma_G$. For better
understanding, we consider the following framework. Let $P$ be a half-plane in $\R^3$ with boundary
the straight line $L$. Put $m_i=\gamma_i\cap P$, $1\leq i\leq 2$. Now, choose any regular curve,
$\alpha:[t_1,t_2]\rightarrow P$, with $\alpha(t_i)=m_i$ and $\alpha^{\prime}(t_i)=\nu (m_i)$,
$1\leq i\leq 2$, and denote by $M_{\alpha}$ the surface of revolution, in $\R^3$, obtained when
rotate $\alpha$ around $L$. The immersion $\phi\in I_{\Gamma}^D(M,\R^3)$ such that
$\phi(M)=M_{\alpha}$ obviously lies in $\Sigma_G$. Conversely, if $\phi\in\Sigma_G$, then we can
regard its image, $\phi(M)$, as a surface of revolution in $\R^3$ obtained when rotate a certain
curve, with the obvious first order boundary data, in $P$ around the axis $L$. Hence, the space
$\Sigma_G$ can be identified with the following class of revolution surfaces

\[ \Sigma_G \equiv \{M_{\alpha} \, / \, \alpha:[t_1,t_2]\rightarrow P, \quad \alpha(t_i)=m_i, \, \alpha^{\prime}(t_i)=\nu(m_i), \,  1\leq i\leq 2\}. \]

\noindent We have proved that the NSM is equivalent to the Willmore one, under any boundary
conditions. This, in particular holds for boundary conditions with rotational symmetry. On the
other hand, the Willmore action is obviously $G$-invariant. Hence, both problems are also
equivalent when reduced, via $G$, to the space of symmetric immersions, $\Sigma_G$. Then, we need
to characterize the critical points of ${\cal W}:\Sigma_G\rightarrow\R$

\vspace{2 mm}

\noindent {\bf Using conformal invariance to reduce variables.} The next idea is to exploit the
extrinsic conformal invariance of the Willmore model with boundary. We take $L$ to be the $z$-axis
and choose $P$ as the right-half-plane in the $\{x,z\}$-plane. The Euclidean space $(\R^3-L,g)$ can
be viewed as the warped product, $P\times_h\S^1$, here $P$ is endowed with its Euclidean metric,
$g_o$ and the warping function $h:P\rightarrow\R$ measures the distance to the axis, $L$. To
characterize the critical points in the NSM with rotational symmetry in the boundary, we use the
conformal invariance in the Willmore variational one. Therefore, in $(\R^3-L,g)$, we make the
conformal change with conformal factor $\frac{1}{h^2}$, that is,
$\bar{g}=\frac{1}{h^2}g=\frac{1}{h^2}g_o+dt^2$. Then, $(\R^3-L,\bar{g})$ is the Riemannian product
of the hyperbolic plane, $(P,\frac{1}{h^2}g_o)$ and the unit circle, $\S^1$. We denote with
overbars the corresponding objects in the new metric, for example $M_{\alpha}$ is a surface of
revolution with boundary $(\R^3-L,g)$ while $\bar{M}_{\alpha}$ is a tube (a Riemannian product)
with boundary in $(\R^3-L,\bar{g})$. Then

\[ {\cal W}(M_{\alpha})=\int_{M_{\alpha}}H_{\alpha}^2dA_{\alpha}+\int_{\partial M_{\alpha}}\kappa^{\alpha} ds=\bar{\cal W}(\bar{M}_{\alpha})=\int_{\bar{M}_{\alpha}}\left(\bar{H}_{\alpha}^2+\bar{R}_{\alpha}\right) d\bar{A}_{\alpha}+\int_{\partial\bar{M}_{\alpha}}\bar{\kappa}^{\alpha}d\bar{s}, \]
where $\bar{R}_{\alpha}$ denotes the sectional curvature of $\bar{g}$ along the surface
$\bar{M}_{\alpha}$ (notice that the corresponding term in the original metric vanishes identically
because $g$ is Euclidean and so flat) and $\bar{\kappa}^{\alpha}$ is the curvature of
$\partial\bar{M}_{\alpha}$ in $\bar{M}_{\alpha}$.

\noindent

Now, $\bar{R}_{\alpha}=0$ because it is a part of the mixed sectional curvature in a Riemannian
product and this vanishes identically. On the other hand, the parallels of tubes are geodesics and
so $\bar{\kappa}^{\alpha}=0$. Therefore, we have

\[ {\cal W}(M_{\alpha})=\int_{M_{\alpha}}H_{\alpha}^2dA_{\alpha}+\int_{\partial M_{\alpha}}\kappa^{\alpha} ds=\bar{\cal W}(\bar{M}_{\alpha})=\int_{\bar{M}_{\alpha}}\bar{H}_{\alpha}^2 d\bar{A}_{\alpha}=\frac{\pi}{2}\int_{\alpha}\bar{\kappa}^2 d\bar{s}, \]

where $\bar{\kappa}$ is nothing but the curvature function of $\alpha$ in the hyperbolic plane $P$.
Notice that to obtain the last equality we have used that parallels are also geodesics in
$(\R^3-L,\bar{g})$.

\vspace{2 mm}

\noindent

Hence, the NSM with rotational symmetry in the boundary is reduced to that for clamped elastica in
the hyperbolic plane. To be precise, in the hyperbolic plane, $(P,\frac{1}{h^2}g_o)$ we choose two
points, $m_1$ and $m_2$, unit vectors $\nu_i (m_i)\in T_{m_{i}}P$ and the space of clamped curves,
$\Lambda=\{\alpha:[t_1,t_2]\rightarrow P, \, \alpha(t_i)=m_i, \, \alpha^{\prime}(t_i)=\nu_i (m_i),
\, 1\leq i\leq 2\}$, and then the variational problem associated with the total elastic energy,
${\cal E}:\Lambda\rightarrow\R$, given by

\[ {\cal E}(\alpha)=\int_{\alpha}\bar{\kappa}^2ds. \]

\vspace{2 mm}

\noindent

{\bf Theorem 2.} {\it The solitons, in the NSM with rotational symmetry in the boundary, that
preserve this symmetry, correspond with the surfaces of revolution obtained by rotation of clamped
elastic curves in the hyperbolic plane, $P$. That is curves that are solutions of the field
equations associated with the boundary valued problem ${\cal E}:\Lambda\rightarrow\R$}.

\vspace{2 mm}

\noindent

{\bf Corollary.} {\it The solitons in the NSM with rotational symmetry in the boundary carry
topological charges that can be determined, holographically from the boundary conditions, to be
$2\pi(\sin{\theta_1}+\sin{\theta_2})$.}

\vspace{2 mm}

\noindent

{\bf Proof.} With the choice $\xi=(0,0,1)$ and $P$ as the right-half-plane in the $\{x,z\}$-plane,
we have

\[ \gamma_i (s)-p_i=(r_i\cos{\frac{s}{r_i}},r_i\sin{\frac{s}{r_i}},0). \]

\noindent

Now, from $\cos{\theta_i}=<\nu_i(s),\xi>$, we obtain

\[ \nu_i (s)=(\sin{\theta_i}\cos{\frac{s}{r_i}},\sin{\theta_i}\sin{\frac{s}{r_i}},\cos{\theta_i}).\]

\noindent

In particular, we have $m_i=(r_i,0,0)$, $\nu_i(m_i)=(\sin{\theta_i},0,\cos{\theta_i})$ and so the
curvature of the parallel $\gamma_i$ in $M_{\alpha}$ is
$\kappa_i^{\alpha}=-\frac{\sin{\theta_i}}{r_i}$. Consequently the topological charge carried by
$M_{\alpha}$ is

\[ Q(M_{\alpha})=\int_{M_{\alpha}}G_{\alpha} dA_{\alpha}=-\sum_{i=1}^{2}\int_{\gamma_i}\kappa_i^{\alpha}(s)ds=2\pi(\sin{\theta_1}+\sin{\theta_2}). \]

\section{Clamped elasticae in Hyperbolic plane}

In this section we conclude the algorithm by describing the moduli space of clamped elastic curves
in the hyperbolic plane.

In $(P,\frac{1}{h^2}g_o)$, assumed to have Gaussian curvature $-1$, we consider the space of
clamped curves, $\Lambda=\{\alpha:[t_1,t_2]\rightarrow P, \, \alpha(t_i)=m_i, \,
\alpha^{\prime}(t_i)=\nu(m_i), \, 1\leq i\leq 2\}$, and the action ${\cal E}:\Lambda\rightarrow\R$,
given by

\[ {\cal E}(\alpha)=\int_{\alpha}\kappa^2ds. \]

\noindent

We use the standard terminology (see \cite{Langer-Singer1} for details) to obtain the following
first variation formula for ${\cal E}$

\[ \delta{\cal E}(\alpha)[W]=\int_{\alpha}<\Omega(\alpha),W>ds+[{\cal R}(\alpha,W)]_{t_{1}}^{t_{2}}, \]

where $\Omega(\alpha)$ and ${\cal R}(\alpha,W)$ stand for the Euler-Lagrange and Boundary
operators, respectively, and they are given by

\[ \Omega(\alpha)=2\nabla^3_T T+3\nabla_T\kappa^2 T-2\nabla_T T, \]

\[ {\cal R}(\alpha,W)=2<\nabla_T W,\nabla_T T>-<W,\nabla^2_T T+3\kappa^2 T>, \]

where $\nabla$ denotes the Levi-Civita connection of the metric $<,>=\frac{1}{h^2}g_o$ in $P$, $T$
is unit tangent vector field of $\alpha$ and $W\in T_{\alpha}\Lambda$.

\vspace{2 mm}

\noindent

On the other hand, we can make the following computations along a curve, $\bar{\alpha}$, in
$\Lambda$ with first order data $(\alpha,W)$

\[ W=d\bar{\alpha}(\partial_r),  \qquad  \nabla_T W=fT+d\bar{\alpha}(\partial_r T), \]

where $f=\partial_r(\log{\mid V\mid})$. Then, we evaluate these formulas along the curve $\alpha$
by making $r=0$ and use the first order boundary data to obtains the following values at the
endpoints

\[ W(t_i)=0,  \qquad  \nabla_T W(t_i=f(t_i)\nu(m_i), \qquad  1\leq i\leq 2. \]

\noindent

As a consequence, the boundary operator drops out, $[{\cal R}(\alpha,W)]_{t_{1}}^{t_{2}}=0$.

\vspace{2 mm}

\noindent

Then, $\alpha\in\Lambda$ is a critical point of the variational problem ${\cal
E}:\Lambda\rightarrow\R$ if and only if $\Omega(\alpha)=0$ and it happens if and only if the
curvature function of $\alpha$ is a solution of the following second order differential equation

\[ 2\frac{d^2\kappa}{ds^2}+\kappa(\kappa^2-2). \]

\noindent

These curves will be called clamped elasticae in the hyperbolic plane and we will briefly describe
them using the free boundary case which was given in \cite{Langer-Singer1}. First, notice that this
equation admit a couple of constant solutions, geodesics ($\kappa=0$) and geodesic circles with
$\kappa=2$. The former, when rotate the limiting case, provides a twice punctured round sphere
which can be completed to obtain the sphere as solution to the $O(3)$-model, while the later gives
an anchor-ring with ratio $\sqrt{2}$.

\vspace{2 mm}

\noindent

When searching for non constant solutions, observe that the equation admits a first integral, which
after the change of variable $u=\kappa^2$ can be written as follows

\[ (u^{\prime})^2=P(u), \qquad P(u)=-u(u^2-4u-4A), \]

where $u^{\prime}=\frac{du}{ds}$. Moreover, a non constant solution, $u=\kappa^2$, must take on
values at which $P(u)>0$. Thus, $P(u)$ has three real roots, say $-a_1,a_2,a_3$, which satisfy
$-a_1\leq a_2\leq a_3$. Now, the general solution of the equation is expressed in terms of elliptic
functions

\[ u=u(s)=a_3(1-q^2 {\rm sn}^2(rs,p)), \]

where

\[ p^2=\frac{a_3-a_2}{a_3+a_1}, \qquad q^2=\frac{a_3-a_2}{a_3}, \qquad r=\frac{\sqrt{a_3+a_1}}{2}. \]

\vspace{2 mm}

\noindent

We distinguish the following possibilities:

\vspace{2 mm}

\noindent

{\bf (A)} If $-a_1=0<a_2<a_3$, then $0<p=q=\frac{\sqrt{a_3-a_2}}{\sqrt{a_3}}<1$ and
$r=\frac{\sqrt{a_3}}{2}$ and then $u(s)=\kappa^2(s)=a_3 {\rm dn}^2(rs,p)$. An elastica with this
curvature function is said to be orbitlike. The qualitative behaviour of these elasticae was
obtained in \cite{Langer-Singer1}. They oscillate between two concentric geodesic circles. Every
piece of an orbitlike elastica gives a clamped elastica which, when rotate, provides a solution to
the NSM. In particular, the class of orbitlike elasticae admits a rational one-parameter subclass
of closed elasticae. They provide, by rotation, genus one compact solutions of the free NSM,
\cite{Langer-Singer2}.

\vspace{2 mm}

\noindent

{\bf (B)} If $-a_1=a_2=0<a_3$, then $p=q=1$ and $r=1$ because the Gaussian curvature is $-1$.
Therefore, the elliptic function providing the curvature becomes into a hyperbolic one, i.e.
$\kappa(s)=2 {\rm sech}(s)$. This elastica is called asymptotically geodesic, it never closes and
it has an integral horocycle. Each piece of an asymptotically geodesic elastica gives a clamped
elastica which provides a surface of revolution being a solution to the NSM.

\vspace{2 mm}

\noindent

{\bf (C)} If $-a_1<a_2=0<a_3$, then $\frac{\sqrt{2}}{2}<p<1$, $q^2=1$ and
$r=\frac{1}{2}\sqrt{a_3+a_1}$. In this case the curvature is $\kappa=\sqrt{a_3} {\rm cn}(rs,p)$ and
the elastica is said to be wawelike. A wavelike elastica can never be closed, however, it
oscillates along an axial geodesic. Pieces of wavelike elasticae give clamped elasticae which
through rotation provide solutions to the NSM. In particular, in the limit, they give twice
punctured genus zero surfaces of revolution which are new solutions to the free NSM.  They can be
completed to get Zoll surfaces with a couple of antipodal singularities.

\section{Conclusions}

In this note, we have developed a geometric algorithm to obtain the whole moduli space of
rotational solitons in the NSM with rotational symmetry in the boundary. This criterion reduces the
search of those solitons to that of clamped elastic curves in a hyperbolic plane.

The main ideas in this method involve the principle of symmetric criticality, the Gauss-Bonnet
formula, the extrinsic conformal invariance of the NSM with boundary and the theory of elasticae in
a hyperbolic plane.

The algorithm works as follows: We choose the boundary axis of symmetry, $L$, as the boundary of
the hyperbolic plane, $P$, regarded as the Poincare half plane model with Gaussian curvature $-1$.
Let $\omega:I\subset\R\rightarrow\R$, defined to be one of the following functions

\begin{eqnarray*} \omega(s) & = & 2a \, {\rm dn}(as,p), \quad a\in\R, \quad {\rm and} \quad 0<p<1, \quad {\rm or}  \\
\omega(s) & = & 2 \, {\rm sech}(s), \quad {\rm or}  \\
\omega(s) & = & a \, {\rm cn}(rs,p), \quad a,r\in\R, \quad {\rm and} \quad \sqrt{2}/2<p<1.
\end{eqnarray*}

\noindent

We choose $\alpha:I\subset\R\rightarrow P$ to be a curve with curvature function $\omega$. For any
$[s_1,s_2]\subset I$, we put $\alpha(s_i)=m_i$, $\alpha^{\prime}(s_i)=v_i$ and $\theta_i$ to denote
the angle that $v_i$ makes with the axis $L$, $1\leq i\leq 2$. In this context, let
$M_{s_1}^{s_2}(\alpha)$ be the surface of revolution obtained when rotate $\alpha([s_1,s_2])$
around $L$. It is obvious that its boundary is $\partial
M_{s_1}^{s_2}(\alpha)=\Gamma=\{\gamma_1,\gamma_2\}$, $\gamma_i$ being the parallel pictured by
$m_i$. It is evident too that the tangent plane $T_p (M_{s_1}^{s_2}(\alpha))$, $p\in\gamma_i$,
makes an angle $\theta_i$ with $L$, $1\leq i\leq 2$. Then, {\bf the Gauss map of
$M_{s_1}^{s_2}(\alpha)$ is a soliton of the NSM with boundary $(\Gamma,\theta_1,\theta_2)$.
Moreover all the solitons preserving the boundary rotational symmetry are obtained in this way. The
charges of these solitons are computed to be $2\pi(\sin{\theta_1}+\sin{\theta_2})$}.

As limiting cases we also obtain solitons of the free NSM when rotate the whole elastica defined in
$\R$. Even if the elastica closes, we get compact solitons as in \cite{Langer-Singer2}.

Finally, it should be noticed that the NSM with boundary is invariant under conformal changes of
the surrounding gravitational field because it is equivalent to the Willmore problem with boundary.
This fact can be used to construct other solitons with different symmetry. For example one can
obtain Hopf tubes with boundary in the three sphere by lifting clamped elasticae in the two sphere
and then to project them, via a suitable stereographic map, to obtain solitons in the NSM with
boundary. The details of this construction so as other related ideas will be developed in a
forthcoming paper.

\vspace{4 mm}

\noindent

{\bf Acknowledgements}

\noindent

This work was partially supported by MCYT Grant BFM2001-2871 with FEDER funds.


\begin{thebibliography}{77}





\bibitem{Cecilia1} C.Albertsson, U.Lindstrom and M.Zabzine, {\em $N=1$ supersymmetric sigma model with boundaries, I.} Comm.Math.Phys., to appear (hep-th/0111161).



\bibitem{Cecilia2} C.Albertsson, U.Lindstrom and M.Zabzine, {\em $N=1$ supersymmetric sigma model with boundaries, II.} preprint (hep-th/0202069).



\bibitem{Anzellotti1} G.Anzellotti, R.Serapioni and I.Tamanini, Indiana Univ.Math.J. 39 (1990), 617.



\bibitem{Anzellotti2} G.Anzellotti and S.Delladio, Proceedings of a Conference in Honor of the 70th birthday of Robert Finn, Stanford University. International Press Incorporated, Boston, Cambridge MA,1995.



\bibitem{Arroyo-Barros-Garay} J.Arroyo, M.Barros and O.J.Garay, Pacific J.Math. 188 (1999), 201.



\bibitem{Arroyo-Barros-Garay1} J.Arroyo, M.Barros and O.J.Garay, J.Phys. A:Math.Gen. 35 (2002), 6815.



\bibitem{Barros} M.Barros, Math.Proc.Camb.Phil.Soc. 121 (1997), 321.



\bibitem{Belavin-Polyakov} A.A.Belavin and A.M.Polyakov, JETP Lett. 22 (1975), 245.



\bibitem{Biscari} P.Biscari, F.Bisi and R.Rosso, J.Math.Biol. 45 (2002), 37.



\bibitem{Cavalcante} F.S.A.Cavalcante, M.S.Cunha and C.A.S.Almeida, Phys.Lett. B 475 (2000), 315.



\bibitem{Concus-Lancaster} P.Concus and K.Lancaster, {\em Advances in Geometric Analysis and Continuum Mechanics.} Stanford University. International Press Incorporated, 1995.



\bibitem{DoCarmo-Dajzer} M.P.Do Carmo and M.Dajzer, T\^{o}hoku Math.J. 34 (1982), 425.



\bibitem{Langer-Singer1} J.Langer and D.A.Singer, J.Diff.Geom. 20 (1984), 1.



\bibitem{Langer-Singer2} J.Langer and D.A.Singer, Bull. London Math.Soc. 16 (1984), 531.



\bibitem{Lawson} H.B.Lawson, Ann. of Math. 92 (1970), 335.



\bibitem{Laughlin} R.B.Laughlin, Phys.Rev.Lett. 60 (1988), 2677.



\bibitem{Ody} M.S.Ody and L.H.Ryder, Int.J.Mod.Phys. A 10 (1995), 337.



\bibitem{Palais} R.S.Palais, Commun.Math.Phys. 69 (1979), 19.



\bibitem{Pinkall} U.Pinkall, {\em Hopf tori in $\S^3$.} Invent.Math. 81 (1985), 379.



\bibitem{Purkait-Ray} S.Purkait and D.Ray, Phys.Lett. A 116 (1986), 247.



\bibitem{Tsurumaru} T.Tsurumaru and I.Tsutsui, Phys.Lett. B 460 (1999) 94.



\bibitem{Weiner} J.L.Weiner, Indiana Univ.Math.J. 27 (1978), 19.



\end{thebibliography}
\end{document}